\documentstyle[a4,12pt]{article}

\def \x{{\bf x}}
\def \y{{\bf y}}
\def \be{\begin{equation}}
\def \ee{\end{equation}}

\def \ra{\rangle}
\def \la{\langle}
\def \1{{\bf 1}}
\def \a{\alpha}
\def \b{\beta}

\def \Tr{\mbox{Tr}}
\def \fr{\frac}

\newfont{\st}{cmr7}

\begin{document}

\begin{titlepage}

\begin{flushright}
DAMTP-93-42\\
\today
\end{flushright}

\vspace*{5mm}

\begin{center}
{\Huge Radial Excited States for Heavy Quark Systems in NRQCD}\\[15mm]
{\large\it UKQCD Collaboration}\\[3mm]

{\bf S.M.~Catterall, F.R.~Devlin, I.T.~Drummond,
 R.R.~Horgan}\\
DAMTP, Silver St., University of Cambridge, Cambridge CB3~9EW, UK

\end{center}
\vspace{5mm}

\begin{abstract}

Following the Non-Relativistic QCD approach we use a gauge invariant
smearing method with factorization to measure the
excitation energies for a heavy $Q\bar{Q}$ system on a $24^3\times 48$
lattice at $\beta=6.2$. The results come from averaging over an ensemble of 60
QCD configurations. In order to enhance the signal from each configuration
we  use wall sources for quark propagators. The quark Hamiltonian contains only
the simplest non-relativistic kinetic energy term.

The results are listed
for a range of bare quark masses. The mass splittings are insensitive
to this variable though there are a slight trends with increasing quark mass.
For an appropriate choice of UV cut-off ($a^{-1}=3.2$Gev)
the mass spectrum compares reasonably well with the experimental
values for the spin-averaged energy gaps of the $\Upsilon$ system.

We also present results for the $DE$ and $DT$ waves for the lowest bare quark
mass.
The results are consistent with degeneracy between the two types of $D$ wave.
This encourages the idea that even with our simple quark Hamiltonian the
departure
from rotational invariance is not great.

\end{abstract}
\vfill
\end{titlepage}

\section{Introduction}

In a previous paper \cite{camb} we studied heavy quark bound states appropriate
 to a
description of the $J/\psi$ and$ \Upsilon$ systems using the non-relativistic
approach of Lepage (NRQCD) \cite{GPL,beth,beth2,GPLetal}. We investigated the
lowest bound states for $S$, $P$ and $D$
waves ignoring spin effects for the quarks using gauge configurations
from the UKQCD collaboration on a
$16^3\times 48$-lattice with a $\b$-value of 6.2. In the present paper we
report
results on the first excitations in the $S$ and $P$-channels for this system
(again without spin). We obtained these results using 60 quenched QCD
configurations from
the UKQCD collaboration on a $24^3\times 48$ lattice at $\b=6.2$~. The new
results are
reasonably consistent with our previous ones but considerably more precise.

Our results are based on the construction of a number of smeared and unsmeared
operators that couple to the appropriate channels and the measurement of their
cross correlators. The smeared operators are constructed in a gauge invariant
manner.
Using a simple subtraction procedure we show that the correlation functions do
indeed
have a multi-exponential structure.
Our best estimates
of the lowest states and the first excited states in both the $S$ and
$P$-channels
of the $Q\bar{Q}$ system are established by performing consistent correlated
fits to the measured operator correlators using appropriately factorizing
two-exponential forms. Some three-exponential fits were attempted to test the
range of applicability of the fits but did not lead to different conclusions.

\section{Quark Propagators}
The quark propagator in a given gauge field background is
\be
G(x,y)=\la \psi(x)\psi^{\dag}(y)\ra~~,
\ee
where the angle brackets indicate averaging over the quark degrees of
freedom $\{\psi(x)\}$
and $x=(\x,t)$, $y=(\y,0)$~. When $t=0$, $G(x,y)=\delta_{\x\y}~$.

The evolution for the (non-relativistic) quark propagator $G(x,y)$ is
\be
G(x+\hat{t},y)=U^{\dagger}_{\hat{t}}(x)\left(1-\frac{H_0}{n}\right)^nG(x,y)
+\delta_{\x\y}\delta_{t0}
\ee
where $\hat{t}$ denotes a unit step in the time direction
and $n$ is the order of the time-step update as discussed by Thacker and
Davies \cite{chris1}.
We set $G(x,y)=0$ for $t\le 0$~.
The modified update is necessary for stability at certain values of the bare
quark mass.
In this paper we use $n=3$~.
The hamiltonian $H_0$ is that appropriate to non-relativistic propagation
 ignoring spin
of a quark of mass $M$, namely
\begin{equation}
H_0={-1\over 2Ma}\sum_{\hat{\mu}=1}^3
\Delta^{+}_{\hat{\mu}}\Delta^{-}_{\hat{\mu}}
\end{equation}
The covariant finite differences $\Delta^{+}$, $\Delta^{-}$ are given by
their usual expressions (we have suppressed all colour and spin
indices)
\be
\Delta^+_{\hat{\mu}}G(x,y)=U_{\hat{\mu}}(x)G(x+\hat{\mu},y)-G(x,y)~~,
\ee
\be
\Delta^-_{\hat{\mu}}G(x,y)=G(x,y)-U^{\dagger}_{\hat{\mu}}(x-\hat{\mu})
G(x-\hat{\mu},y)~~.
\ee

\section{Smeared Operators}
The ideal method for detecting excitated states in a given channel
is to construct operators each of which couples only to one of the states.
An alternative approach is to accept as a starting point a basis set of
operators
with quantum numbers appropriate to the channel of interest and to recognize
that an intermediate state will couple to each of these operators in a unique
 way so that
the exponential contribution associated with that state to the cross
correlators of the
basis operators will have a factorizing form. This is the approach we have
adopted in dealing
with the gauge invariant smeared operators we construct and use in our
simulation.

The operators we investigated for the $S$-channel were in addition to the
standard
point operator (we use $\chi(x)$ to denote the anti-quark degrees of freedom)
\be
O^{(0)}(x)=\chi^{\dag}(x)\psi(x)+\mbox{h.c.}~~,
\ee
a set of operators of the form
\be
O^{(m)}(x)=\sum_{\hat{\mu}}\chi^{\dag}(x)\left(M^{m}_{\hat{\mu}}(x)
\psi(x+m\hat{\mu})
                         + \hat{M}^{m}_{\hat{\mu}}(x)\psi(x-m\hat{\mu})\right)
                                            +\mbox{h.c.}~~,
\ee
where the $\hat{\mu}$-sum is over space like directions and the matrices
 $M^{m}_{\mu}(x)$ and $\hat{M}^{m}_{\mu}(x)$
have the (appropriately ordered) product forms
\be
M^{m}_{\hat{\mu}}(x)=\prod_{\nu=0}^{m-1}
U_{\hat{\mu}}(x+\nu\hat{\mu})~~\mbox{and}~~
\hat{M}^{m}_{\hat{\mu}}(x)=\prod_{\nu=1}^{m}
U^{\dag}_{\hat{\mu}}(x-\nu\hat{\mu})~~.
\ee

For the $P$-channel we use a family of operators of the form
\be
O_{\hat{\mu}}^{(m)}(x)=\chi^{\dag}(x)\left(M_{\hat{\mu}}^{m}(x)
\psi(x+m\hat{\mu})
                        -\hat{M}_{\hat{\mu}}^{m}(x)\psi(x-m\hat{\mu})\right)
                                     +\mbox{h.c.}~~,
\ee

The $DE$ wave operators are
$$
O_{\hat{\mu}\hat{\nu}}^{DE(m)}(x)=\chi^{\dag}(x)\left(M_{\hat{\mu}}^{m}(x)
\psi(x+m\hat{\mu})
          +\hat{M}_{\hat{\mu}}^{m}(x)\psi(x-m\hat{\mu})\right.~~~~~~~~~~~~~~~~
$$
\be
            ~~~~ ~~~~~~~~~~~~~~~~ \left.-M_{\hat{\nu}}^{m}(x)\psi(x+m\hat{\nu})
                        -\hat{M}_{\hat{\nu}}^{m}(x)\psi(x-m\hat{\nu})\right)
                                     +\mbox{h.c.}~~,
\ee
and the $DT$ wave operators are
\be
O_{\hat{\mu}\hat{\nu}}^{DT(m)}(x)=\chi^{\dag}(x)\left(\Delta_{\hat{\mu}}^{(m)}
\Delta_{\hat{\nu}}^{(m)}
                      +\Delta_{\hat{\nu}}^{(m)}\Delta_{\hat{\mu}}^{(m)}\right)
\psi(x)+~\hbox{h.c.}~~,
\ee
where
\be
\Delta_{\hat{\mu}}^{(m)}\psi(x)=M_{\hat{\mu}}^{m}(x)
\psi(x+m\hat{\mu})-\hat{M}_{\hat{\mu}}^{m}(x)\psi(x-m\hat{\mu})~~.
\ee

The correlation functions we measure are
\be
F^{S}_{nm}(t)=\fr{1}{V}\sum_{\x,\y}\la O^{(n)}(x)O^{(m)}(y)\ra~~,
\label{gr1}
\ee
for $S$-wave analysis, and
\be
F^{P}_{nm}(t)=\fr{1}{3V}\sum_{\hat{\mu},\x,\y}\la O_{\hat{\mu}}^{n}(x)
O_{\hat{\mu}}^{m}(y)\ra~~,
\label{gr2}
\ee
for the $P$-wave analysis. Here $V=24^3$, the spatial lattice volume.

All of the above operators are of the form
\be
O(x)=\chi^{\dag}(x)\Psi(x)~~,
\ee
where for an appropriate set of ($SU(3)$-matrix) coefficients $\{C_{\x\x'}\}$
\be
\Psi(x)=\sum_{\x'}C_{\x\x'}\psi(x')~~,
\ee
and $x'=(\x',t)$~.
A typical correlation function can be expressed as
\be
F_{12}(t)=\fr{1}{V}\sum_{\x\y}\la \Tr \bar{G}_{12}(x,y)
G^{\dag}(x,y)\ra+\mbox{c.c.}~~,
\label{gr3}
\ee
where
\be
\bar{G}_{12}(x,y)=\la \Psi^{(1)}(x)\Psi^{(2)\dag}(y)\ra~~,
\label{gr4}
\ee
where 1 and 2 indicate the two (possibly the same) smeared operators.
Of course the Green's function $\bar{G}_{12}(x,y)=\sum_{\x'}
C^{(1)}_{\x\x'}{\cal{G}}^{(2)}(x',y)$
where ${\cal{G}}^{(2)}(x',y)$ can be calculated with  an appropriate
change of initial condition by the same method as the original quark
Green's function.

\section{Wall Source Method}

Because the computing overhead is considerable it is desirable to extract
 as much signal
as possible from each pass through a gauge field configuration. To this end
we modify the
 measured correlators $F^{S}_{nm}(t)$, $F^{P}_{nm}(t)$ and $F^{D}_{nm}(t)$as
follows. Eq(\ref{gr3})
leads to a method of computation for our typical correlation
function $F_{12}(t)$
that requires the evaluation of the Green's functions for at least a
 representative sample
of $y$-values on the initial time slice if we wish to maximize the information
to be extracted
from each configuration. This is computationally onerous.
An alternative procedure is the following.
We replace the single $\y$-summation in eq(\ref{gr3}) with a double sum thus
\be
F_{12}(t)=\fr{1}{V}\sum_{\x,\y,\y'}\la\Tr \bar{G}_{12}(x,y)
G^{\dag}(x,y')\ra+\mbox{c.c.}~~,
\ee
where $y'=(0,\y')$~. We now rely on the gauge field averaging to eliminate the
contributions from the gauge non-invariant off-diagonal terms in the above
double $(\y,\y')$-sum
leaving only the contribution from the gauge invariant diagonal terms for
which $\y=\y'$~.
The disadvantage of the method is that the off-diagonal contributions
provide noise
even if they do average to zero. The advantage of the method is that we pick
 up all the
diagonal terms in one pass since it is only necessary to compute the objects
of the form
\be
g(x)=\sum_{\y}G(x,y)~~,
\ee
which satisfies the same equation as $G(x,y)$ and the initial condition
\be
g(0,\x)=1~~.
\ee
Similar remarks apply to $\bar{g}_{12}(x)=\sum_{\y}\bar{G}_{12}(x,y)$~.
We have then
\be
F_{12}(t)=\fr{1}{V}\sum_{\x}\la\Tr \bar{g_{12}}(x)g^{\dag}(x)\ra~~,
\ee
In practice we do find that the method works well and does provide a good
signal
relatively economically. It is implicit in the discussion that no gauge fixing
has been imposed on the ensemble of gauge fields. However because of the
limitations
of the data set the averaging procedure may not work perfectly. The different
 treatment
of the two operators in the correlation
function may mean that the symmetry $F^{S,P,D}_{mn}(t)=F^{S,P,D}_{nm}(t)$ no
longer holds. This
does not destroy factorization and we allow for the asymmetry in fitting
 the data.
This procedure is similar to the multiple origin approach first
utilised by
Kenway\cite{kenway} and Billoire et al.\cite{bill1}, except we seed all
 sites on the initial timeslice.
A quark wall source was also used by Gupta et al.\cite{gupta1} although
 they fix to
Coulomb gauge.

\section{Data Fitting}

As indicated above we try to fit the correlation function $F^{S}_{mn}(t)$
with the
multi-exponential form
\be
F^{S}_{mn}(t)=\sum_{a}\gamma_{(a)m}\gamma'_{(a)n}e^{-M_{(a)}t}~~.
\label{exp1}
\ee
Our main results are obtained using a two exponential form requiring eight
 parameters.
This allows us to obtain estimates for the lowest $S$ and $P$ states together
 with the
first excitations. Our statistical method involved a correlated least squares
fit based
on an estimate of the complete set of variances and cross correlators of all
the fitted
quantities. Since our results are based on 60 statistically independent gauge
field
configurations we felt that eight parameters was a reasonable number for the
 fitted form.
We did perform three exponential fits in certain cases with twelve parameters.
Where these
seemed reliable they were consistent with the two exponential fits but with
considerably less tight errorbars. The results we quote are from separate $S$
and $P$-channels fits.
We also carried out a combined $S$ and $P$-channel fit but obtained results
that were
little different.

The limitations of the data led us to confine ourselves to two operators per
channel in any one fit.
The precise
form of the $2\times2$ matrix of correlators was
$$
\left(\begin{array}{cc}F_{mm}(t)&F_{mn}(t)\\
                 F_{nm}(t)&F_{nn}(t)\end{array}\right)~~~~~~~~~~~~~~~~~~~~~
    ~~~~~~~~~~~~~~~~~~~~~~ ~~~~~~~~~~~~~~~~~~~~~~~~~~~~~~~~~~~~~~~~
$$
\be
{}~~~~~=
  \left(\begin{array}{cc}(\gamma^{(1)}_m)^2&\gamma^{(1)}_m
\gamma^{(1)}_n\rho^{(1)}\\
               \gamma^{(1)}_n\gamma^{(1)}_m&(\gamma^{(1)}_n)^2\rho^{(1)}
\end{array}\right)e^{-Mt}+
\left(\begin{array}{cc}(\gamma^{(2)}_m)^2&\gamma^{(2)}_m\gamma^{(2)}_n
\rho^{(2)}\\
               \gamma^{(2)}_n\gamma^{(2)}_m&(\gamma^{(2)}_n)^2\rho^{(2)}
\end{array}\right)e^{-(M+\Delta M)t}~~.
\ee
This is equivalent to the form in eq(\ref{exp1}). The asymmetry in the
factorized forms is
represented by the departure of the parameters $\rho^{(1)}$ and $\rho^{(2)}$
 from unity.
Note that we have parametrized the splitting $\Delta M$ between the two
levels explicitly
since this is the quantity of direct interest.

The basis of the fitting procedure is the estimate of the correlation matrix of
results. At any one time these comprised the two direct and two cross
correlators
for two operators evaluated on 48 time slices. The correlation matrix was
therefore
of dimension $192\times 192$~. Our data is extracted from 60 independent gauge
 configurations.
The correlation matrix is therefore of rank $r\le 60$~ and therefore
necessarily singular.
In practice the effective rank of the correlation matrix is even less than
this since
beyond a certain point the eigenvalues become so small their estimation from
 the data is not
reliable. The least squares fitting procedure and the associated error
estimates require the use
of the inverse of the correlation matrix. It is necessary and indeed correct
 to restrict
the inversion of the matrix to an appropriate subspace that is spanned by
eigenvectors
with eigenvalues large enough for reliable estimation from the data.
The dimension of the subspace is referred to as the Singular Value
Decomposition (SVD) cut.

In assessing the results of the fitting procedure we examined cases with a
range of values
of initial off-set and SVD cuts for different combinations of smeared
operators. Our criterion
for a choice of result was that the $\chi^2$-value be acceptably near unity
 per degree of
freedom and that the error was the best (usually the first) of a range of
reasonably good and statistically
consistent fits.

\section{ Explicit Diagonalization Scheme}

Before exhibiting the results of the correlated fits we show directly the
 existence
of a second exponential by means of a  diagonalization method.
L\"uscher and Wolff  \cite{LuWol} have
shown that the eigenvalues of the correlation matrix are of the form
\be
e^{-M_{(a)}t} \left( 1 + O(e^{-\Delta M_{(a)}t}) \right)~~
\ee
where $\Delta M_{(a)}$ is the distance of state $M_{(a)}$ from other states.
  Thus we
evaluate the eigenvalues of the correlation matrix using the appropriate
Numerical
Recipes routines \cite{NuRe}.
Fig. 1 shows the result for $Ma=1.5$ for the $S$-wave combination
$O^{S}_0(x)$ and $O^{S}_4(x)$~ - the ground state
$S$-wave is suppressed revealing the existence of the exponential associated
 with the
first excited state.
Fig. 2 shows effective mass plots for the $1S$ and $2S$ states
obtained from these graphs. Figs. 3~\&~4 show similar results for the $1P$
and $2P$ states.
The results are reasonably consistent with those of the correlated fits
discussed below
which were used to produce the quoted numbers.

In order to obtain reasonably smooth plots the effective mass was defined
as
\be
M(t)=0.25*\log\left(\fr{A(t)}{A(t+4)}\right)~~,
\ee
where
\be
A(t)=\left(F(t)+wF(t+1)+w^2F(t+4)+w^3F(t+3)\right)/4~~,
\ee
and $w$ is chosen to render the terms in the sum of comparable size.

\section{Correlated Fits for S and P Waves}

The results of the correlated fits are shown in Table 1~. Also shown are the
operators used to obtain the results, the $\chi^2$ per degree of freedom, the
SVD-cuts and the off-sets at which a reasonable statistical stability set in.

\begin{table}[h]
\centering
\begin{tabular}{|c|c|l|l|l|c|c|c|}\hline
$M_0a$&Chan.&$Ma$&$\Delta Ma$&$\chi^2$/dof&SVD-cut&Off-set&Operators m\\ \hline
1.5 & $S$& 1.1152(8)&0.198(24)&25.6/18&26&17&0 and 4\\ \cline{2-8}
    & $P$& 1.243(12)&0.116(12)&19.2/20&28&11&2 and 6\\ \cline{1-8}
2.0 & $S$& 0.9788(7)&0.194(12) &11.53/12&20&12&1 and 4\\ \cline{2-8}
    & $P$& 1.112(13)&0.105(20)&14.48/20&28&11&2 and 6\\ \cline{1-8}
3.0 & $S$& 0.8287(7)&0.179(17)&21.68/18&26&15&0 and 4\\ \cline{2-8}
    & $P$& 0.969(17)&0.090(17)&11.6/14&22&11&2 and 6\\  \hline
\end{tabular}
\caption{The results for the ground states and first excited splits
in the $S$ and the $P$ channels. }
\end{table}

It is also interesting to compare the various mass splits with the
spin-averaged
values of the $\Upsilon$-system. We use a conversion factor from lattice units
to physical units of $a^{-1}=3.2$~. The results are listed in Table 2~.

\pagebreak

It is clear that the pattern of mass splitting for the lower two
bare masses ($M_0a=1.5~\&~2.0$) is reasonably close to the actual splitting for
the $\Upsilon$ except for the $2S$ level which appears to be too high but
exhibits a downward trend as $M_0a$ increases. The other splits are less
sensitive to
changes in the bare mass.

\begin{table}[h]
\centering
\begin{tabular}{|c|l|l|l|l|l|}\hline
$M_0a$&$1S-2S$ &$1S-1P$  &$1S-2P$  &$1P-2P$\\ \hline
1.5  &0.198(24)&0.128(12)&0.244(17)&0.116(12)\\
     &0.634(77) Gev&0.410(38) Gev&0.781(54) Gev&0.371(38) Gev  \\ \hline
2.0  &0.194(8) &0.133(13)&0.238(24)&0.105(20) \\
     &0.621(26) Gev&0.426(42) Gev&0.762(77) Gev&0.336(64) Gev \\ \hline
3.0  &0.179(17)&0.140(17)&0.230(24)&0.090(17) \\
     &0.573(54) Gev&0.448(54) Gev&0.736(77) Gev&0.288(54) Gev \\ \hline\hline
Expt($\Upsilon$)&0.563 Gev&0.430 Gev&0.795 Gev&0.365 Gev \\\hline\hline
\end{tabular}
\caption{The results for various mass splits in lattice and physical units
compared to the spin-averaged results for the $\Upsilon$-system. The
conversion factor
is $a^{-1}=3.2$ Gev. }
\end{table}

The ratios of mass splits is independent of the choice
for $a^{-1}$~. For each bare quark mass these ratios are listed in Table 4
taking the central value of the $1S-1P$ split as the base.

\begin{table}[h]
\centering
\begin{tabular}{|c|l|l|l|l|l|}\hline
$M_0a$&$1S-2S$ &$1S-1P$  &$1S-2P$  &$1P-2P$\\ \hline
1.5  &1.55(19)&1.00(12)&1.91(13)&0.91(9)\\ \hline
2.0  &1.46(6) &1.00(13)&1.79(18)&.79(15) \\ \hline
3.0  &1.28(12)&1.00(17)&1.64(17)&0.64(12) \\ \hline
Expt($\Upsilon$)&1.31&1.00&1.85&0.85 \\\hline\hline
\end{tabular}
\caption{The results for various mass splits expressed as ratios to the central
value of the $1S-1P$ split and
compared to the spin-averaged results for the $\Upsilon$-system. }
\end{table}
These results are not dissimilar to those of the phenomenological
 non-relativistic
quark models \cite{Lucha}. The spectrum is relatively independent of the bare
 quark mass
though the $2S$ state is rather high.
Apart from the anomalously high $2S$ state the ratios that fit best
correspond to a bare
quark mass somewhere between $M_0a=1.5$ and $M_0a=2.0$~. Using the same
conversion factor as
above we find these correspond to $M_0=4.8$ Gev and $M_0=6.4$ Gev. This is
to be compared with
the $B$-quark mass of $\sim 5$ Gev suggested by the mass of the $\Upsilon$
 itself.
If we take the $1S-1P$ mass split for $M_0 a=1.5$ as
the correct basis on which to calculate then we
find $a^{-1}=3.4(3)$ Gev which encompasses the above value.

Our results for $a^{-1}$ are higher than suggested by a measurement of the
string tension $\sigma$~.
At $\b=6.2$, $\sigma a^2=0.026(1)$ \cite{alltonetal,bali} that is
 ${\sqrt\sigma}a=0.161(3)$~. If we use
the phenomenological value ${\sqrt\sigma}=.42$ Gev
we obtain $a^{-1}=2.6(1)$ Gev. This is in line with other estimates of
$a^{-1}$.
Another way of expressing this discrepancy is to note that our lattice
calculation
at $\b=6.2$ yields a ratio ${\sqrt\sigma}/\Delta M(1S-1P)\simeq 1.25$ whereas
 the phenomenological
result is $\simeq 1.0$~. The question then is whether or not there is a
reasonable
explanation of this discrepancy for our model. One answer is to recognize that
 the
string tension is associated with the long range part of the quark potential
 while the
$1S-1P$ split comes about as a result of a balance between long and short range
effects in the potential. The short range force is controlled by the strong
coupling
evaluated at a higher momentum, $q^*$, than that, $\bar{q}$, associated with
 the string tension.
The main difference between
the quenched and unquenched theories is the differential renormalization of
the strong
coupling $\a(q)$ at a given $q$ due to the vacuum polarization effects of
light quarks.
If we fix the string tension to be the same in both theories then we have
$\a_U({\bar q})=\a_Q({\bar q})$
However because the quenched coupling runs faster than the unquenched one we
 have $\a_U(q^*)>\a_Q(q^*)$~.
The short range force in the quenched theory will therefore be weaker than
in the unquenched case.
Because the $S$-wave states are particularly sensitive to the short range
part of the $Q-\bar Q$ force
they will be more deeply bound in the unquenched theory than in the quenched
one.
The $P$-waves will be controlled more by the longer range part of the force
associated with the
string tension. This will tend to leave the $P$-waves unchanged between the
 two theories
with the result that for a given string tension $\Delta M(1S-1P)$ will be
greater in the
unquenched relative to the quenched theory. In turn this will yield a lower
 value for
the ratio $\sqrt{\sigma}/\Delta M(1S-1P)$ for the unquenched relative to the
 quenched
theory in line with our results. Similar results will hold for for any two
quantities
associated with different momentum scales. In the quenched theory they will
yield
different estimates for $a^{-1}$ while unquenched theory (by definition) will
 produce
consistent estimates.

\section{Correlated Fits for D Waves}

In Table 4 we show the results for a two exponential fit to the two versions
of the
$D$ wave operators for the bare quark mass $M_0a=1.5$~.

\begin{table}[h]
\centering
\begin{tabular}{|c|c|l|l|l|c|c|c|}\hline
$M_0a$&Chan.&$Ma$&$\Delta Ma$&$\chi^2$/dof&SVD-cut&Off-set&Operators m\\ \hline
1.5 & $DE$& 1.373(11)&0.315(13)&23.7/18&26&6&2 and 6\\ \cline{2-8}
    & $DT$& 1.388(19)&0.389(41)&22.5/14&22&6&1 and 4\\ \cline{1-8}
\end{tabular}
\caption{The results for the ground states and first excited splits
in the $DE$ and the $DT$ channels. }
\end{table}
The encouraging feature of these results is the degeneracy within errors not
only of the basic states in the two channels but also of the first excited
states.
The conclusion is that to a good approximation the cubical symmetry of the
(spatial) lattice
is replaced by rotatational symmetry. Expressed in physical units
 $\Delta M(1S-1D)=0.83(4)$ Gev
if we again use $a^{-1}=3.2$ Gev.
There is so far no observed $D$ wave for the $\Upsilon$ system but the
corresponding
state for charmonium ($J/\psi$) has $\Delta M(1S-1D)=0.702$ Gev. Given the
simple and
approximate nature of our heavy quark Hamilton this is an encouraging result.
The quality of the results from the simulation restricted the application of
the
fitting procedure to the range $t<20$ so the outcome for the mass gap may be
expected
to be on the high side. This circumstance may also explain why the
measured gap $\Delta M(1D-2D)\simeq$ 1 Gev is implausibly high. It is
interesting that it showed
in both versions of the $D$ wave spectrum.

\section{Conclusions}

We have measured the $Q\bar{Q}$ mass splittings for the radial excitations of
the $S$ and $P$
waves using the non-relativistic heavy quark propagators calculated from
quenched gluon
configurations from the UKQCD collaboration. Our measurements were based
on 60 QCD
configurations on a $24^3\times 48$ lattice. The Hamiltonian we used to
calculate the
quark propagators was of the simplest kind containing only the kinetic energy
 contribution
and omitting all higher corrections. The results emerged from two-exponential
correlated fits to pairs of smeared operator correlators.
Our best results yielded statistical errors $\sim 10\%$ for the mass
splittings.

On a broad picture our results are not inconsistent with the pattern of
 spin-averaged
splittings of the $\Upsilon$-system. In particular the ratios
$\Delta M(1S-1P):\Delta M(1S-2P):\Delta M(1S-1D)$
seem roughly correct. The $2P$ wave shows a slight dependence on the bare
quark mass. We have not yet
determined the dependence of the $1D$ wave on this mass. The absence of an
experimental $1D$ state for
$\Upsilon$ restricts us to a comparison with the corresponding charmonium
state or theoretical quark
model predictions but that comparison is encouraging. If we base our
evaluation of the cut-off strictly
on the $1S-1P$ mass split then we find the value $a^{-1}=3.4$ Gev. This
very much in line with scaling
predictions from results of the corresponding calculations performed at
 values of $\b=$5.7 and 6.0 \cite {LepagePriv}~.
This suggested a bare quark mass for the kinetic energy Hamiltonian in the
 range 4.8 to 6.4 Gev.

However there are two obvious problems that present themselves.
The first is that the ratio involving the string tension
$\sqrt{\sigma}/\Delta M(1S-1P)$ is measured as 1.25 compared to a
 phenomenological value of 1.0~. It is
plausible that this discrepancy may be due to the overly strong running of
the coupling in the quenched theory.
The second is the anomalously high value of $\Delta M(1S-2S)/\Delta M(1S-1P)$
and its sensitivity to
the bare quark mass. It may be that the deficiencies of the quenched
approximation can resolve this problem
also. An alternative explanation is that there is a measurement problem
with the $S$ wave channel.
Future measurements using different smeared operators constructed
in the Coulomb gauge from quark model wave functions should help to resolve
 the issue.

\section*{Acknowledgements}

We thank G P Lepage and C Davies for enlightening discussions.
F R Devlin wishes to acknowledge the support of the Department of Education
of Northern Ireland.

\vfill

\pagebreak

\vfill
\pagebreak

\section*{Figure Captions}
\begin{enumerate}

\item[] Fig. 1 The results for the $S$-wave propagator $F^{(S)}_{00}(t)$
(circles) together with the
the results of the diagonalization procedure applied to the correlation matrix
 for the
operators $O^{(S)}_{0}$ and $O^{(S)}_{4}$ (diamonds)
revealing the contribution of the $2S$-state.

\item[] Fig. 2 The effective mass plot for the $1S$ (upper graph) and $2S$
 states (lower graph).
The estimates obtained from the correlated fit are indicated by a full line.

\item[] Fig. 3 The results for the $P$-wave propagator $F^{(P)}_{22}(t)$
(circles) together with the
the results of the diagonalization procedure applied to the correlation matrix
 for the
operators $O^{(P)}_{2}$ and $O^{(P)}_{6}$ (diamonds)
revealing the contribution of the $2P$-state.

\item[] Fig. 4 The effective mass plot for the $1P$ (upper graph) and $2P$
states (lower graph).
The estimates obtained from the correlated fit are indicated by a full line.

\end{enumerate}

\end{document}